\documentstyle[11pt,mrs2001,epsfig]{article}
\begin{document}
\title{Mass Profiles of Clusters at Large Radii from Weak Gravitational Lensing}

\author{D. Clowe $^1$, P. Schneider $^1$}
\affil{$^1$IAEF, Universit\"at Bonn, Auf dem H\"ugel 71, 53121 Bonn, Germany}

\begin{abstract}
We present the weak lensing mass reconstructions of three clusters of galaxies
from data taken on the ESO Wide Field Imager.  We detect a lensing signal in
all three clusters over a range of $1\arcmin < r < 17\arcmin$ ($\approx
0.1 h^{-1} \mathrm{Mpc} < r < 2 h^{-1} \mathrm{Mpc}$).  We measure the best
fitting SIS and NFW profiles for each cluster and show that in two of the
three clusters the NFW profile is preferred to the SIS with marginal
significance ($\sim 2\sigma$).
\end{abstract}

\section{Introduction}
In the past few years, high-resolution N-body simulations have been used to 
make predictions of the shapes of mass profiles for clusters of galaxies
(e.g.~\cite{NFW}, hereafter NFW).  These profiles are characterized with a
shallower, although still singular, core ($\rho \propto r^{-1}$ inside 100 kpc
for NFW) than the classical singular isothermal sphere (SIS), but a steeper 
profile at large radii ($\rho \propto r^{-3}$ outside 1 Mpc).  While the
deviation at small radii is currently being investigated with X-ray and strong
gravitational lensing studies (e.g.~\cite{Schmidt01}), a perhaps more promising
method to distinguish which of the profile classes is correct is measuring
the mass profiles of the clusters at large radii with weak gravitational
lensing.

Weak lensing is a study in which one measures the ellipticities of background
galaxies and looks for a statistical deviation from an isotropic ellipticity
distribution.  This has the advantages over other methods to measure cluster
masses that one gets a direct measure of the mass with no assumptions
regarding they dynamical state of the cluster, and that, with current
technology, the maximum radius at which one can measure the signal is limited
primarily by the size of the detectors.  To date most weak lensing cluster
observations are limited to the central $h^{-1}$ Mpc (e.g.~\cite{clowe00}),
but with the advent of large mosaic CCD cameras, we can now measure a weak
lensing signal in half-degree fields with relatively little telescope time.

\begin{figure}
\plotone{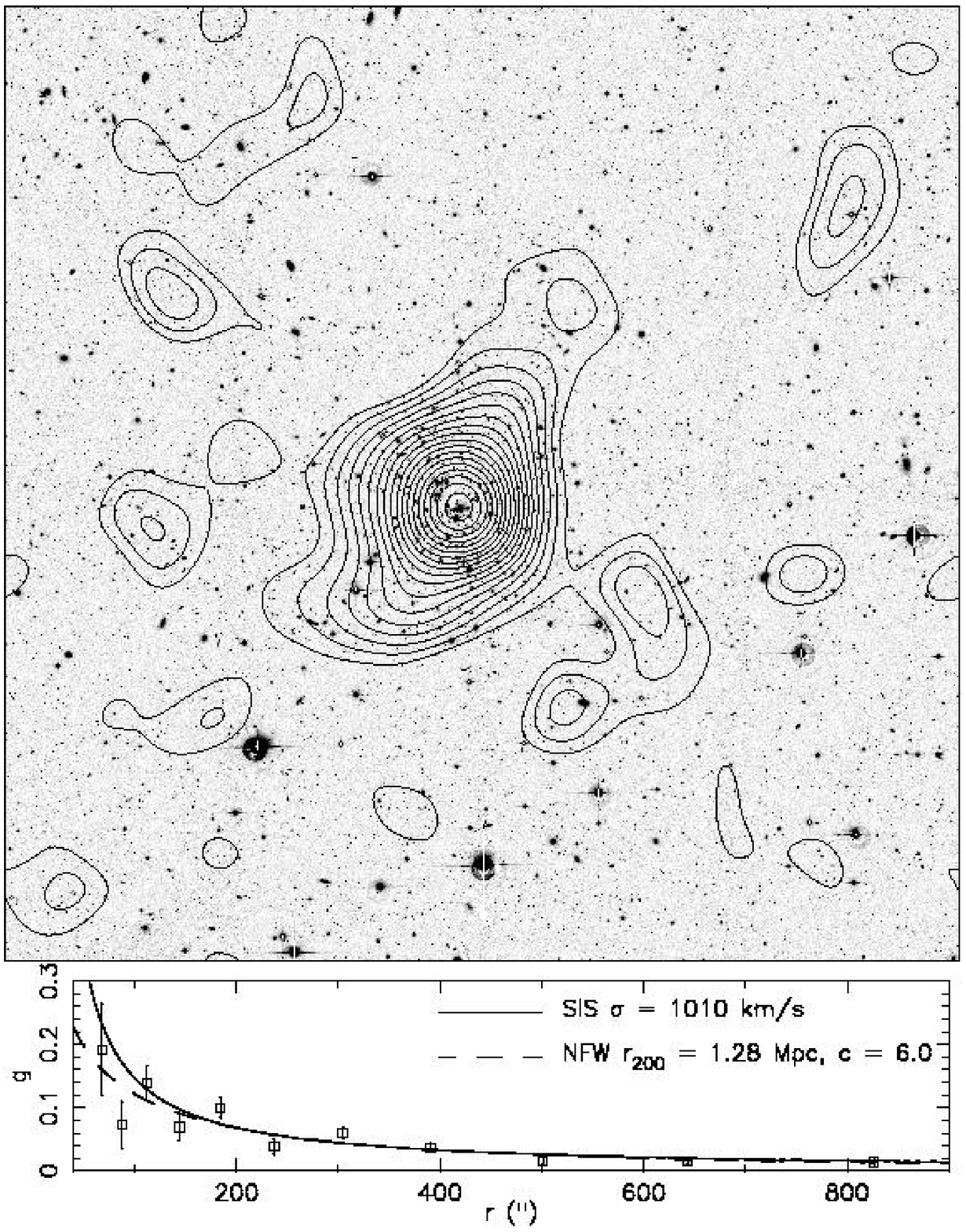}{6.3in}
\caption{Shown above is the WFI R-band image of A1689 ($32\farcm 8$, 
$\sim 4.28 h^{-1} \mathrm{Mpc}$, on a side)
with the weak lensing mass reconstruction overlayed in contours.  The mass
reconstruction has been smoothed by a $1\farcm 15$ Gaussian and each contour
represents a change in $\kappa $ of 0.01 above the mean at the edge of the
image.  Plotted below the image is the azimuthally averaged reduced shear
profile, centered on the brightest cluster galaxy, in logarithmically spaced
bins in radius.  Also plotted are the best fitting SIS and NFW profiles.  No
correction factors have been made for the presence of foreground and cluster
dwarf galaxies and faint stars in the background galaxy sample used to 
measure the shear.  Correcting for this would increase the mass of the
best fit models by 20-30$\%$ \cite{clowe01}.}
\end{figure}

\begin{figure}
\plotone{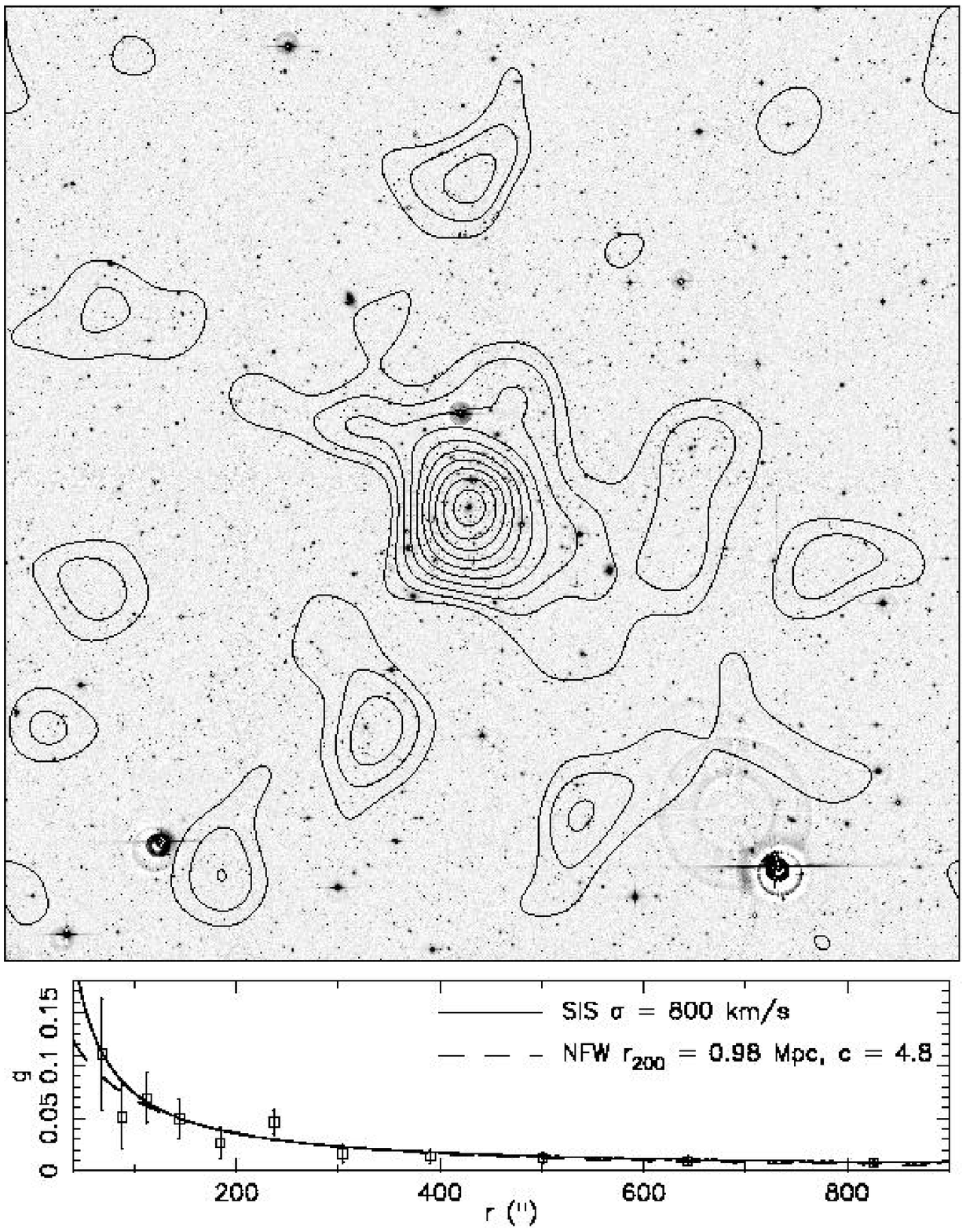}{6.3in}
\caption{Shown above is the WFI R-band image of A1835 ($32\farcm 8$, $\sim 5.42 h^{-1} \mathrm{Mpc}$, on a side)
with the weak lensing mass reconstruction overlayed in contours.  The mass
reconstruction has been smoothed by a $1\farcm 15$ Gaussian and each contour
represents a change in $\kappa $ of 0.01 above the mean at the edge of the
image.  Plotted below the image is the azimuthally averaged reduced shear
profile, centered on the brightest cluster galaxy, in logarithmically spaced
bins in radius.  Also plotted are the best fitting SIS and NFW profiles.  No
correction factors have been made for the presence of foreground and cluster
dwarf galaxies and faint stars in the background galaxy sample used to 
measure the shear.  Correcting for this would increase the mass of the
best fit models by 20-30$\%$.}
\end{figure}

\begin{figure}
\plotone{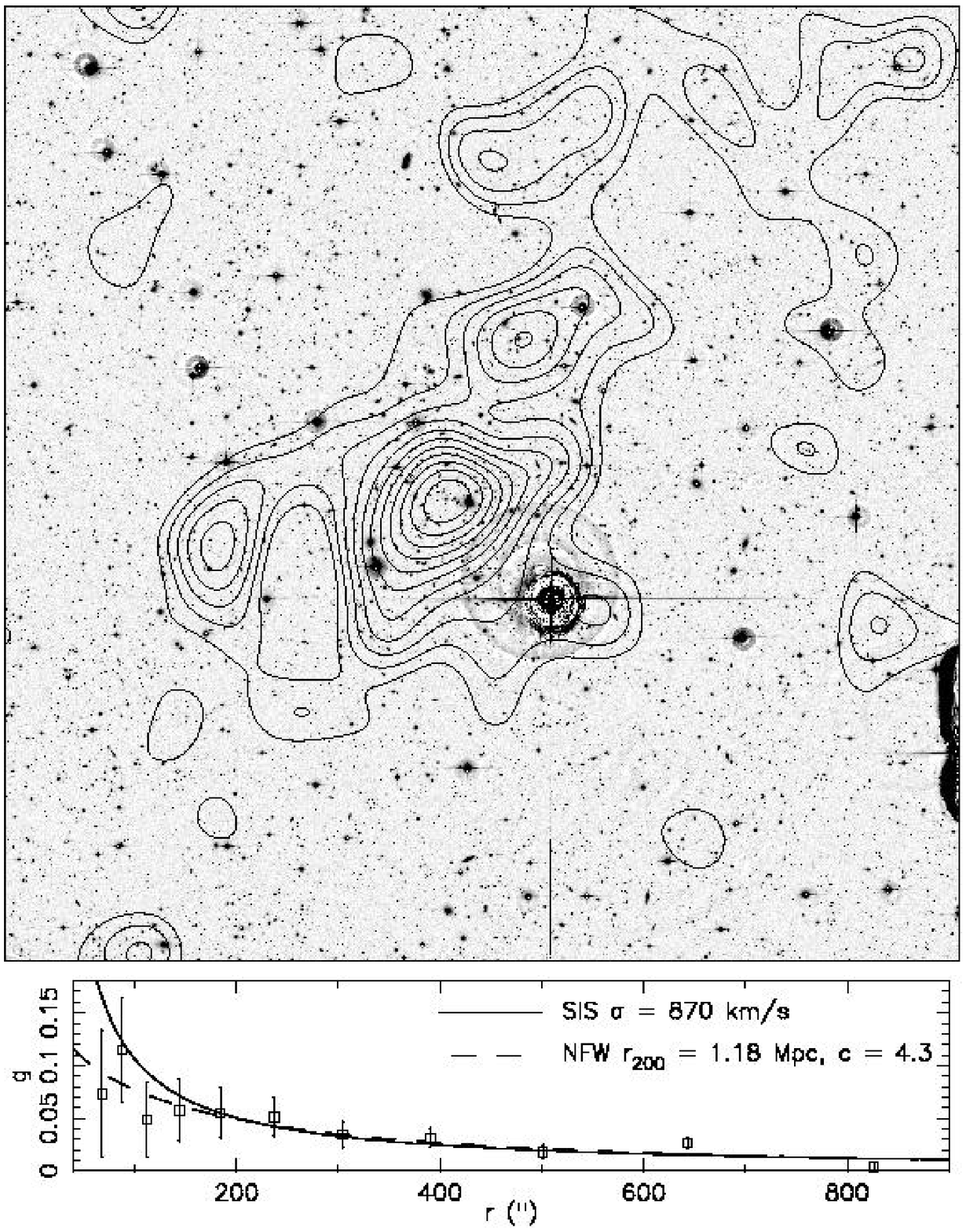}{6.3in}
\caption{Shown above is the WFI R-band image of A2204 ($32\farcm 8$, $\sim 3.63 h^{-1} \mathrm{Mpc}$, on a side)
with the weak lensing mass reconstruction overlayed in contours.  The mass
reconstruction has been smoothed by a $1\farcm 15$ Gaussian and each contour
represents a change in $\kappa $ of 0.01 above the mean at the edge of the
image.  Plotted below the image is the azimuthally averaged reduced shear
profile, centered on the brightest cluster galaxy, in logarithmically spaced
bins in radius.  Also plotted are the best fitting SIS and NFW profiles.  No
correction factors have been made for the presence of foreground and cluster
dwarf galaxies and faint stars in the background galaxy sample used to 
measure the shear.  Correcting for this would increase the mass of the
best fit models by 20-50$\%$.}
\end{figure}

\section{Observations and Data Reduction}
We observed the clusters A1689 ($z=0.182$), A1835 ($z=0.252$), A2204 
($z=0.152$), and A1347 on the nights of May 29 and 30, 2000 using the
Wide Field Imager at the ESO/MPG 2.2m on La Silla.  Each cluster was 
observed with 12 900s exposures in the R-band, which resulted in coadded
images of $\sim 35\arcmin \times 34\arcmin$, with $\sim 87\%$ of each
image having the full 3 hours of integration, and the remaining $13\%$ having
some smaller amount as the region was out of the field of view or between
the gaps in the chip mosaic during some of the exposures.  The final images
have a FWHM of bright but unsaturated stars of $\sim 0\farcs 7$.  
Object counts at a $3\sigma$ detection limit in the regions containing the
full exposure time are complete to $R = 24.9$ for $2\arcsec $ radius aperture
magnitudes, as measured by the point where the number counts depart from a
power law.

Objects were detected and had their sizes, magnitudes, and second moments of
the surface brightness measured using the IMCAT software package written by
Nick Kaiser.  Ellipticities were created from the measured second moments
and were corrected for PSF anisotropy and circular smearing using the KSB
techniques \cite{KSB}.  The corrected ellipticities can then be used
as a direct estimator of the reduced shear, $g$, at the location of the galaxy.
A1347 appears to consist of several small, unrelated groups of galaxies and
stars clusters, and as such was used as a blank field to test that the observed
weak lensing signals are in fact caused by the mass of the clusters and not
by camera or telescope distortions.  

The resulting cluster shears were then analyzed with two methods.  The first
method was using an inversion algorithm which uses the fact that both the
shear and the mass surface density are second derivates of the surface 
potential to create a two dimensional image of the mass \cite{KS93}.
This surface density distribution, however, can only be determined to within
an unknown additive constant.
The second method bins the shear radially around a chosen center of mass, the
brightest cluster galaxy in this case.  The resulting shear profile can then
be fit with various mass profiles.  The results for both techniques are shown
for A1689, A1835, and A2204 in Figures 1, 2, and 3 respectively.

\begin{table}
\begin{center}
\caption{best fit mass models}
\begin{tabular}{cccccccccc}
\hline
Cluster & \multicolumn{3}{c}{SIS model} & \multicolumn{4}{c}{NFW model} &
\multicolumn{2}{c}{F-test} \\
&$\sigma$ (km/s)&$\chi ^2/274$&S/N&$r_{200}$ (Mpc)&$c$&$\chi ^2/273$&S/N&
F&Prob.\\\hline
A1689&1030&0.921&13.7$\sigma$&1.28&6.0&0.910&13.6$\sigma$&4.18&96$\%$\\
A1835&800&1.095&7.9$\sigma$&0.98&4.8&1.097&7.7$\sigma$&0.46&50$\%$\\
A2204&870&1.060&8.1$\sigma$&1.18&4.3&1.049&8.0$\sigma$&3.54&94$\%$\\
\hline\hline
\end{tabular}
\end{center}
\end{table}

\section{Discussion}
As can be seen in the figures, all three clusters are detected at high 
signal-to-noise and the shear in the outermost bin is detected at greater
than $3\sigma$ in A1689 and A1835, and greater than $3\sigma$ for the 
sum of the two outmost bins in A2204.  
The best fit mass models and their significances are given
in table 1.  All of the clusters have the center of the highest mass peak
spatially coincident, to within the errors in the reconstruction, with the 
position of the brightest cluster galaxy, and all significant mass peaks
have a corresponding peak in the galaxy distribution.  Because the fields
have been studied only in a single color, we do not have any information
regarding which of the additional peaks are substructure within the cluster
and which are foreground or background mass peaks.  As such, while there
appear to be filamentary structures extending from all three clusters
in both the mass reconstructions and galaxy distributions, we cannot be sure
that this is a detection of mass flowing into the cluster.

Also shown in table 1 are the best fit SIS and NFW models to the radial
shear profiles of the clusters.  While all three clusters are well fit by
both profiles, as determined by the reduced $\chi ^2$, an F-test of an
additional term, used to compare $\chi ^2$ values with different degrees of
freedom, gives that the NFW profile is preferred over the SIS
profile in A1689 and A2204 with $96\%$ and $94\%$ confidence.  All of the
cluster mass models, however, are much less massive than those predicted
from the X-ray observations of the clusters 
\cite{clowe01}\cite{Schmidt01},\cite{bcs}.  This is at least partially due 
to having only a single passband
to use to select the background galaxy catalog.  As a result, the catalog of
objects with $23<R<25.5$ used in the mass reconstructions, 
which we assumed were only background galaxies ($z_{\mathrm{bg}} \sim 1$),
is contaminated with faint stars and foreground and cluster dwarf galaxies.
Estimates of the degree of contamination from other fields at similar galactic
latitudes predict that 10-20$\%$ of the objects are faint stars, and a similar
amount are (primarily cluster) dwarf galaxies.  Thus, to correct for this
contamination, the shear measurements and masses of the best fit models would
need to be increased by $\sim$20-50$\%$.  Observation of these fields in
additional passbands will allow for the selection of background galaxies by
color, which should greatly reduce the level of foreground contamination.

\vfill
\end{document}